\documentclass[showkeys,12pt, preprint,preprintnumbers,nofootinbib, groupedaddress,superscriptaddress,amsmath,amssymb]{revtex4}
\usepackage{graphicx}
\usepackage{dcolumn}
\usepackage{bm}
\usepackage{amssymb}
\usepackage{amsmath}
\usepackage{epsfig}    
\usepackage{color}
\usepackage{slashed}
\usepackage{hhline}

\def\be{\begin{equation}}
\def\ee{\end{equation}}
\newcommand{\bea}{\begin{eqnarray}}
\newcommand{\eea}{\end{eqnarray}}
\newcommand{\nn}{\nonumber}

\numberwithin{equation}{section}

\begin{document}

{\begin{flushright}{KIAS-P17132}
\end{flushright}}

\title{Radiative seesaw models linking to dark matter candidates inspired by the DAMPE excess} 
\author{Takaaki Nomura}
\email{nomura@kias.re.kr}
\affiliation{School of Physics, KIAS, Seoul 02455, Korea}

\author{Hiroshi Okada}
\email{macokada3hiroshi@cts.nthu.edu.tw}
\affiliation{Physics Division, National Center for Theoretical Sciences, Hsinchu, Taiwan 300}

\date{\today}

\begin{abstract}
We propose two possibilities to explain an excess of electron/positron flux around 1.4 TeV recently reported by Dark Matter Explore (DAMPE) in the framework of
 radiative seesaw models where one of them provides a fermionic dark matter candidate, and the other one provides a bosonic dark matter candidate.
 We also show unique features of both models regarding neutrino mass structure.

\end{abstract}
\maketitle
\newpage

\section{Introduction}
In light of the excess of electron/positron flux reported by Dark Matter Explore (DAMPE)~\cite{dampe,Ambrosi:2017wek}, 
we try to propose possibilities whether radiative neutrino models can accommodate dark matter (DM) candidate that can explain the excess and link DM to neutrinos or not.
The typical features of DM indicated by the excess are that DM mainly annihilates into electron positron pair which originates from very sharp excess at the distribution energy $\sim$1.4 TeV, and no excess of proton antiproton pair is observed~\cite{Yuan:2017ysv}.
Another important feature is that scale of the annihilation cross section to explain the excess is $3\times 10^{-26}{\rm cm^3/s}$ {if we adopt a scenario which assume an existence of dark subhalo near the earth. Remarkably this scale of the DM annihilation cross section} is similar to the annihilation cross section to explain the relic density. It suggests that we do not need to worry about large boost factor any longer to enhance the cross section, which is unlikely to the previous experimental results of positron excess reported by PAMELA~\cite{Adriani:2008zr} and AMS-02~\cite{Accardo:2014lma}.
{In fact, these positron excess can be addressed by astrophysical sources such as pulsars while it is difficult to explain the excess in the DAMPE data due to the sharpness of the peak.}

In view of model building~\cite{Fan:2017sor, Gu:2017gle, Duan:2017pkq, Zu:2017dzm, Tang:2017lfb, Chao:2017yjg, Gu:2017bdw, Athron:2017drj, Cao:2017ydw, Duan:2017qwj, Liu:2017rgs, Chao-guo-li-shu, Huang:2017egk, Gao:2017pym, Niu:2017hqe}, the simplest  achievement could be to introduce a flavor dependent gauged $U(1)$ symmetry that has to have nonzero electron/positron charge at least; therefore $U(1)_{e-\mu}$ or  $U(1)_{e-\tau}$.
If DM is fermion, Dirac type is favored because its cross section should still have s-wave dominant in the s-channel $\bar XX\to Z'\to \bar ff$,
where $Z'$ is extra gauge boson induced from additional gauged  $U(1)$ symmetry.
In case of bosonic DM candidate, there could be several possibilities to explain the excess due to a lot of modes coming from Higgs potential as well as Yukawa term that depends on models. As an example, in ref.~\cite{Zu:2017dzm, Cao:2017ydw}, they have discussed the possibility via four body electron/positron final states in the s-channel.

In our letter, we firstly construct the one-loop induced neutrino model with Dirac type of DM propagating inside the loop, introducing gauged $U(1)_{e+\mu-\tau}$ symmetry. In order to cancel the gauge anomalies, several mirror fermions have to be introduced, but it does not violate our model once additional discrete symmetry is imposed. As a result, we show not only successful neutrino scenario but also predictive neutrino texture.
In the second model, we consider the bosonic DM candidate propagating inside the loop diagram generating neutrino mass, introducing  gauged $U(1)_{e-\mu}$ symmetry based on ref.~\cite{Lee:2017ekw}. Then we also show its successful features simultaneously explaining the muon anomalous magnetic dipole moment.

This paper is organized as follows. 
In Sec.~II, we explain two models, one of which is a radiative neutrino model with Dirac DM candidate, and another of which is the one with bosonic DM candidate. 
Finally we summarize the results in Sec.~III.

 \section{Model, particle properties and phenomenologies}
\begin{center} 
\begin{table}
\begin{tabular}
{|c||c|c|c|c||c|c||c|c|c|c|c|c|}\hline\hline  
 & \multicolumn{4}{c||}{SM Leptons}  & \multicolumn{2}{c||}{Neutral fermions} & \multicolumn{5}{c|}{Mirror Leptons}\\\hline
Fermions  ~&~ $L_{L_{\ell'}}$  ~&~ $L_{L_\tau}$ ~&~ $\ell'_{R}$ ~&~ $\tau_R$ ~&~ $N_{\ell'}$ ~&~ $N_{\tau}$~&~ $L'_{R(L)_{\ell'}}$ ~&~ $L'_{R(L)_\tau}$ ~&~ $e'_{L(R)}$ ~&~ $\mu'_{L(R)}$ ~&~ $\tau'_{L(R)}$
\\\hline 
 $SU(2)_L$  & $\bm{2}$    & $\bm{2}$   & $\bm{1}$    & $\bm{1}$  & $\bm{1}$    & $\bm{1}$  & $\bm{2}$  & $\bm{2}$    & $\bm{1}$  & $\bm{1}$   & $\bm{1}$\\\hline 
$U(1)_Y$  & $-\frac{1}{2}$  & $-\frac12$  & $-1$   &  $-1$  &  $0$  &  $0$ & $-\frac{1}{2}$ & $-\frac12$  & $-1$ &  $-1$  &  $-1$\\\hline
 $U(1)_{e+\mu-\tau}$ & $1$  & $-1$ & $1$   & $-1$ & $1$    & $-1$  & $1(0)$ & $-1(0)$ & $1(0)$  & $1(0)$   & $-1(0)$  \\\hline
$Z_4$  & $+1$  & $+1$ & $+1$ & $+1$ & $-1$ & $-1$  & $\pm i$  & $\pm i$ & $\pm i$ & $\pm i$ & $\pm i$\\\hline
\end{tabular} 
\caption{Field contents of fermions
and their charge assignments under $SU(2)_L\times U(1)_Y\times  U(1)_{e+ \mu -\tau}\times Z_4$, where $SU(3)_C$ singlet for all fermions and $\ell' = e, \mu$.  Notice that one generation of mirror leptons with $U(1)_{e+ \mu -\tau}$ charge 1 can cancel gauge anomalies. }
\label{tab:1}
\end{table}
\end{center}

\begin{table}[t]
\centering {\fontsize{10}{12}
\begin{tabular}{|c||c|c|c||c|c|}\hline\hline
&\multicolumn{3}{c||}{VEV$\neq 0$} & \multicolumn{2}{c|}{Inert } \\\hline
  Bosons  &~ $\Phi_1$  ~ &~ $\Phi_2$~ &~ $\varphi$ ~&~ $\eta'$   ~ &~ $\eta$ ~ \\\hline
$SU(2)_L$ & $\bm{2}$  & $\bm{2}$ & $1$  & $\frac12$  & $\frac12$   \\\hline 
$U(1)_Y$ & $\frac12$  & $\frac12$ & $0$ & $\frac12$  & $\frac12$    \\\hline
 $U(1)_{e+\mu-\tau}$ & $0$  & $-2$  & $1$  & $-2$ & $0$   \\\hline
$Z_4$ & $+1$ & $+1$ & $+1$  & $-1$ & $-1$ \\\hline
\end{tabular}%
} 
\caption{Field contents of bosons
and their charge assignments under $SU(2)_L\times U(1)_Y\times U(1)_{e+\mu-\tau}\times Z_4$, where $SU(3)_C$ singlet for all bosons. }
\label{tab:2}
\end{table}

{In this section, we introduce our models and discuss some phenomenologies including neutrino mass generation, DM relic density and possibility to explain the DAMPE excess.}

\subsection{Model 1: Dirac fermion DM}
{Here we consider a model that provides Dirac fermion DM based on $U(1)_{e+\mu-\tau}$ gauge symmetry.}
In the fermion sector, we introduce vector-like neutral fermions $N_{{e,\mu,\tau}}$ with isospin singlet and several mirror fermions that are needed to cancel our gauge anomalies, and impose a flavor dependent gauge symmetry  $U(1)_{e+\mu-\tau}$ as summarized in Table~\ref{tab:1}.
 Also a discrete $Z_4$ symmetry is imposed for this new fermion in order to forbid the tree level neutrino masses as well as mixing between mirror fermions and other fermions, and stabilize our DM candidate~\footnote{Another types of models have been proposed by refs.~\cite{Nomura:2017xko, Nomura:2017tzj, Nomura:2017psk, Nomura:2017lsn},  several of which are extended to be quark sector.}.

In the scalar sector, we add an $SU(2)_L$ doublet scalar $\Phi_2$, two $SU(2)_L$ doublet inert scalars $\eta,\ \eta'$ and a singlet scalar $\varphi$ to the SM-like Higgs $\Phi_1$ as summarized in Table~\ref{tab:2}.
Notice here $\Phi_{1,2},\varphi$ have the vacuum expectation values (VEVs) after spontaneous symmetry breaking, which are respectively symbolized by $v_{1,2}/\sqrt2$, $v'/\sqrt2$.
 $Z_4$ odd parity is  imposed to assure inert feature of  $\eta,\ \eta'$.

{\it Yukawa interactions}:
Under these fields and symmetries, the renormalizable relevant Lagrangian for neutrino sector and crucial terms of the Higgs potential is given by 
\begin{align}
-{\cal L}_L &=  f_\alpha \bar L_{L_\alpha} \tilde\eta N_{R_\alpha} 
+ g_i \bar N_{L_i} L^c_{L_i} \tilde\eta' + h_{\alpha\beta} \bar N_{L_\alpha} L^c_{L_\beta} \tilde\eta
 + {M_\alpha} \bar N_{L_\alpha} N_{R_\alpha}  + {\rm h.c.}, \nn\\
V \supset & \frac{\lambda_1} 2(\eta^\dag \Phi_1)^2  + \lambda_2(\Phi_1^\dag \eta)(\Phi_2^\dag \eta') + \lambda_3 (\Phi_1 \Phi_2^\dagger) (\varphi^*)^2  + {\rm h.c.},
\label{eq:lag-quark}
\end{align}
where $\alpha(\beta)$ runs over $(e,\mu,\tau)$, $i$ runs over $(e,\mu)$,
$\tilde \eta\equiv i\sigma_2 \eta^*$, $\sigma_2$ is the second Pauli matrix, and we omit trivial terms in the potential.
{Note that $\lambda_3$ term in the potential prevents Higgs doublet sector from inducing massless Goldstone boson.}
Notice also that $h_{\alpha\beta}$ allows only the nonzero components $h_{e,\tau},h_{\mu,\tau},h_{\tau,e},h_{\tau\mu}$,
and dominant mass terms for mirror fermions are induced via $\bar L'_{R} L'_{L} \varphi$ and $\bar \ell'_L \ell'_R \varphi$ where $\ell=e,\mu,\tau$.~\footnote{They also have Yukawa terms $\bar L'_{R(L)_\ell} \Phi_1 \ell'_{L(R)}$ as a subdominant contributions to their masses.}
{For Yukawa interactions among two Higgs doublets and SM fermions, the second doublet $\Phi_2$ can couple to only leptons as a consequence of extra U(1) charge. In this paper. we omit further discussion. }

We parametrize the scalar fields as 
\begin{align}
&\Phi_{1(2)} =\left[
\begin{array}{c}
w^+\\
\frac{v_{1(2)}+\phi_{1(2)}+iz_{1(2)}}{\sqrt2}
\end{array}\right],\ 
{\eta} =\left[
\begin{array}{c}
\eta^{+}\\
\frac{\eta_R+i\eta_I}{\sqrt2}
\end{array}\right]
,\ 
{\eta'} =\left[
\begin{array}{c}
\eta^{'+}\\
\frac{\eta'_R+i\eta'_I}{\sqrt2}
\end{array}\right],\
\varphi=\frac{v'+\varphi_R+i\varphi_I}{\sqrt2},
\label{component}
\end{align}
where $v=\sqrt{v_1^2+v_2^2}\simeq 246$ GeV is VEV of the Higgs doublets, and $w^\pm$, $z_1$, and $\varphi_I$ are respectively GB 
which are absorbed by the longitudinal component of $W$, $Z$, and $Z'(\equiv Z_{e+\mu-\tau})$ boson.
Then we have the three by three CP-even  mass matrix squared $m^2_{R}$ in the basis of $[\phi_1, \phi_2,\varphi_R]^T$.
This is then diagonalized by $O_{R}^T m^2_R O_{R}\equiv$Diag[$m_{h_1},m_{h_2},m_{h_3}$],
where $h_1\equiv h_{SM}$.
The mass eigenstates of  inert bosons in basis of $[\eta_R,\eta'_R]^T$ and $[\eta_I,\eta'_I]^T$ are defined as
\begin{align}
&\eta_R=c_a H_1+ s_a H_2,\quad  \eta'_R=-s_a H_1+ c_a H_2,\\
&\eta_I=c_b A_1+ s_b A_2, \quad \eta'_I=-s_b A_1+ c_b A_2,
\end{align}
where $c(s)_{a(b)}$ are written in terms of $\lambda_{1,2}$ and trivial parameters in the Higgs potential.

{We have heavy neutral gauge boson $Z'$ after $U(1)_{e+\mu-\tau}$ gauge symmetry breaking.
The mass of $Z'$ is given by
\begin{equation}
m_{Z'} = g' \sqrt{4 v_2^2 + v'^2},
\label{eq:MZp1}
\end{equation}
where $g'$ is the gauge coupling of $U(1)_{e+\mu-\tau}$.}

\begin{figure}[!hptb]
\begin{center}
\includegraphics[width=14cm]{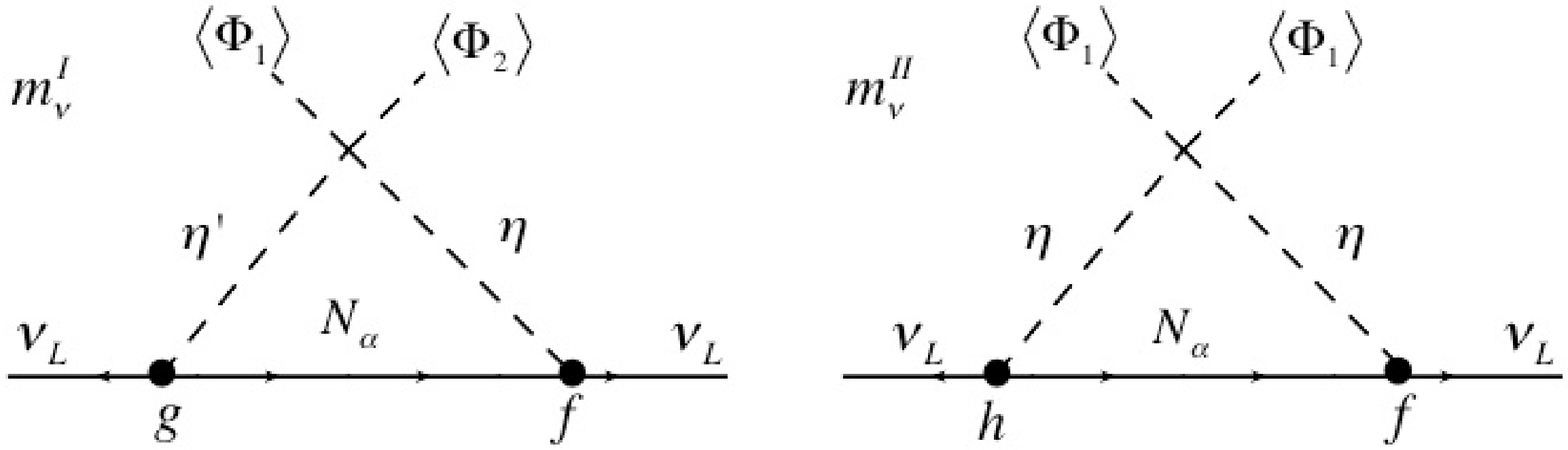}
\caption{The Feynman diagram for generating neutrino mass matrix for Model 1.  } 
  \label{fig:neutrino}
\end{center}\end{figure}

Let us consider the active neutrino mass matrix at one-loop level~\cite{Ma:2006km}, which consists of two types of diagrams; $m^{I}_\nu$ and $m^{II}_\nu$ as shown in Fig.~\ref{fig:neutrino}.
At first, let us consider the relevant Lagrangian for $m^{I}_\nu$ in terms of mass eigenstate:
\begin{align}
{\cal L}&=\frac{f_\alpha}{\sqrt2}\bar\nu_{L_\alpha} N_{R_\alpha} (c_a H_1+s_a H_2)
-i \frac{f_\alpha}{\sqrt2}\bar\nu_{L_\alpha} N_{R_\alpha} (c_b A_1+s_b A_2)\nn\\
&+\frac{g_i}{\sqrt2}\bar N_{L_i} \nu^c_{L_i} (-s_a H_1+c_a H_2)
-i \frac{g_i}{\sqrt2}\bar N_{L_i} \nu^c_{L_i} (-s_b A_1+c_b A_2).
\end{align}
Then 
it is formulated by
\begin{align}
&  (m^{I}_\nu)_{ii} \approx\sum_{i=e,\mu}
 \frac{f_i M_i g_i}{2(4\pi)^2}
 [s_a c_a F_I(M_i,m_{H_1},m_{H_2})-s_b c_b F_I(M_i,m_{A_1},m_{A_2})]
\equiv\left[\begin{array}{ccc}  m_{11} & 0 & 0 \\
0&  m_{22} &  0 \\
0 &  0  & 0 \\ 
 \end{array}\right],\\
&F_I(m_1,m_2, m_3)=\frac{m_1^2 m_2^2\ln\frac{m_1^2}{m_2^2}+m_1^2 m_3^2\ln\frac{m_3^2}{m_1^2}+m_2^2 m_3^2\frac{m_2^2}{m_3^2}}{(m_1^2-m_2^2)(m_1^2-m_3^2)},
\end{align}
where $m_{H(A)_{1,2}}$ is the mass of $H(A)_{1,2}$.
Next, let us consider the relevant Lagrangian for $m^{II}_\nu$ in terms of mass eigenstate:
\begin{align}
{\cal L}&=\frac{f_\alpha}{\sqrt2}\bar\nu_{L_\alpha} N_{R_\alpha} (c_a H_1+s_a H_2)
-i \frac{f_\alpha}{\sqrt2}\bar\nu_{L_\alpha} N_{R_\alpha} (c_b A_1+s_b A_2)\nn\\
&+\frac{h_{\alpha\beta}}{\sqrt2}\bar N_{L_\alpha} \nu^c_{L_\beta} (c_a H_1+s_a H_2)
-i \frac{h_{\alpha\beta}}{\sqrt2}\bar N_{L_\alpha} \nu^c_{L_\beta} (c_b A_1+s_b A_2).
\end{align}
Then it is also formulated by
\begin{align}
&  (m^{II}_\nu)_{\alpha\beta} \approx
-\sum_{\alpha,\beta=e,\mu,\tau}
 \frac{f_\alpha M_\alpha h_{\alpha\beta}}{2(4\pi)^2}
\left( [2 (c_a^2- c^2_b)\ln\Delta_1 + 2 (s_a^2- s^2_b)\ln\Delta_2]\right.\nn\\
&\left.
+\frac{c^2_b m^2_{H_1}-c^2_a m^2_{A_1}}{m_{H_1}^2-m_{A_1}^2} F_I(M_\alpha,m_{H_1},m_{A_1})
+\frac{s^2_b m^2_{H_2}-s^2_a m^2_{A_2}}{m_{H_2}^2-m_{A_2}^2} F_I(M_\alpha,m_{H_2},m_{A_2})\right)
\\
&\hspace{3cm}
\equiv\left[\begin{array}{ccc}  0 & 0 & m_{13} \\
0& 0 &  m_{23} \\
m_{13} &   m_{23}  & 0 \\ 
 \end{array}\right].
\end{align}

The resulting neutrino mass matrix $m_\nu\equiv m^{I}_\nu+m^{II}_\nu$ is finally found to be~\footnote{Another type of model has been proposed by ref.~\cite{Baek:2015mna}.} 
\begin{align}
m_\nu \approx 
\left[\begin{array}{ccc}
 m_{11} & 0 &  m_{13} \\
0& m_{22} &  m_{23} \\
 m_{13}  &  m_{23}  & 0 \\ 
 \end{array}\right].
\end{align}
This is known as type-$B_2$ of two-zero texture that provides several predictions such that  inverted hierarchy is favored when the best fit observables are adapted,
$m_{\nu_3}\approx \sqrt{\Delta m^2_{\rm atm}/(1-\cot^4\theta_{23})}$ and $m_{\nu_2}\approx m_{\nu_1}\approx m_{\nu_3}\cot^2\theta_{23}$ are derived at the leading order~\footnote{See ref.~\cite{Fritzsch:2011qv} in details.}, where $m_{\nu_{1,2,3}}$ and $\Delta m_{\rm atm}$
are respectively observed neutrino mass eigenvalues and atmospheric neutrino mass difference squared~\cite{pdg}.
Here we define $m_\nu=U D^\nu U^T$, $D^\nu\equiv {\rm diag.}[m_{\nu_1} e^{2i\rho},m_{\nu_2} e^{2i\sigma},m_{\nu_3}]$, $U$ is 3 by 3 unitary mixing matrix, and $\rho,\sigma$ are Majorana phases.

{\it Lepton flavor violations(LFVs)} have to always be taken into account in any radiative seesaw models.
The stringent bound arises from the process of $\mu\to e\gamma$; ${\rm BR}(\mu\to e\gamma)\lesssim 4.2\times 10^{-13}$~\cite{TheMEG:2016wtm}.
To satisfy this bound, the typical Yukawa coupling is of the order 0.1 when the loop masses are of the order 1 TeV.
Since we expect the minimal mass inside the loop is about 1.5 TeV that is DM inspired by DAMPE,
this constraint can be applied. Therefore LFVs do not give so serious constraint.

\begin{figure}[t!]
\centering 
\includegraphics[width=8.4cm]{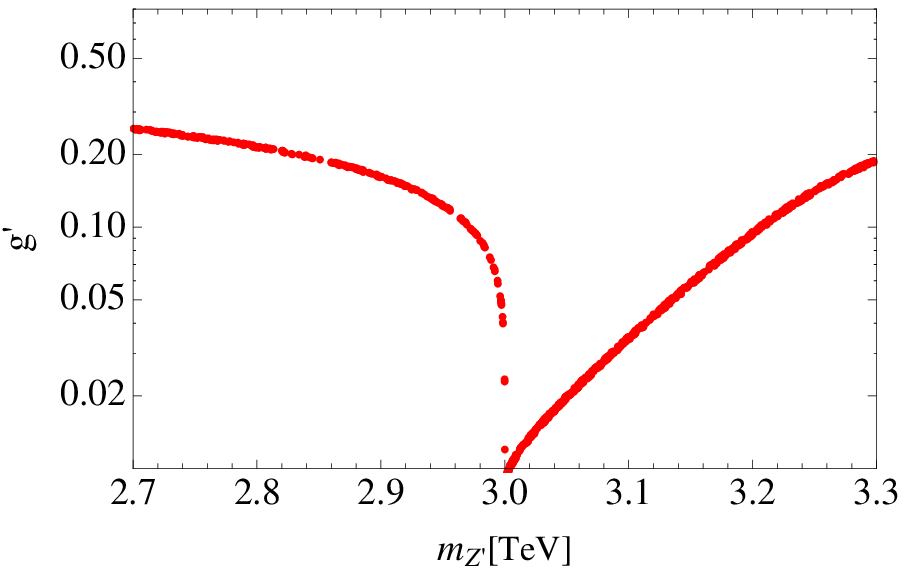}%
\includegraphics[width=8cm]{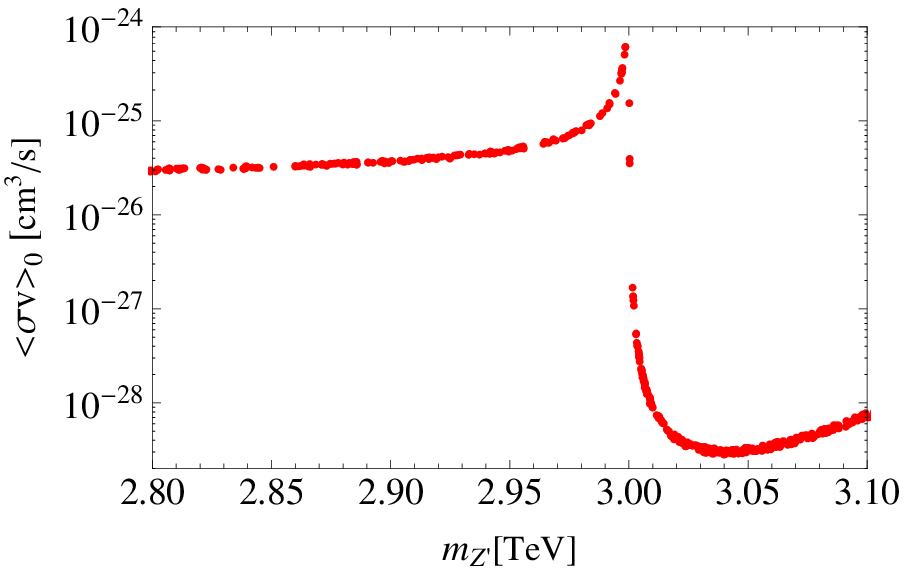}%
\caption{Left plot: The parameter points providing correct relic density on $\{m_{Z'}, g' \}$ plane.  Right plot: Thermally averaged annihilation cross section at current Universe as a function of $m_{Z'}$ using parameter points in the left plot.  Here the DM mass is fixed to be $M_X = 1.5$~TeV.}
\label{fig:DM1}
\end{figure}
{{\it Dark matter candidate inspired by DAMPE}:
Here we discuss possible explanation of the DAMPE data.
At first we briefly formulate the valid interactions between Dirac DM candidate ($X\equiv N_1,\ M_X\equiv M_1$) and the other particles.
Since our DM has extra U(1) charge it comes from kinetic terms including a gauge field, and the interacting Lagrangian is given by
\begin{align}
{\cal L}\sim g'\bar X\gamma^\mu X Z'_\mu 
+\sum_{f=e,\mu} \bar f \gamma^\mu f Z'_\mu+\sum_{f=\nu_e,\nu_\mu} \bar f \gamma^\mu P_Lf Z'_\mu
- \bar \tau \gamma^\mu\tau Z'_\mu- \bar \nu_\tau \gamma^\mu P_L \nu_\tau Z'_\mu,
\end{align}
where the mass of $Z'$ is given by Eq.~(\ref{eq:MZp1}).
Thus $X$ annihilates via the process $\bar X X \to Z' \to \ell^+ \ell^- (\bar \nu_\ell  \nu_\ell)$ and lepton ratio from DM annihilation is $e: \mu: \tau = 1:1:1$ which is also suitable to explain the DAMPE data~\cite{Yuan:2017ysv}.
Here we estimate relic density of $X$ using {\tt micrOMEGAs 4.3.5 }~\cite{Belanger:2014vza} to solve
the Boltzmann equation by implementing relevant interactions which induce DM pair annihilation processes.
In explaining the DAMPE excess, we fix DM mass as $M_X = 1.5$ TeV and scan parameter space $\{g', m_{Z'} \}$ to search for the region providing correct relic density.
In addition, we take into account the constraints from LEP data for measurement of $e^+e^- \to \ell^+ \ell^-$ processes as~\cite{Schael:2013ita} 
\begin{equation}
\frac{m_{Z'} }{g'} \geq 7.0 \ {\rm TeV}
\end{equation}
where upper limit of gauge coupling is $g' \sim 0.4$ for $m_{Z'} \sim 3$ TeV.
The left plot in Fig.~\ref{fig:DM1} shows the parameter points which can accommodate correct DM relic density~\cite{Ade:2015xua}, approximated as $\Omega h^2 = 0.12 \pm 0.005$. 
Then we show the current thermal annihilation cross section for the parameter points in the right plot in Fig.~\ref{fig:DM1}.

Then we can explain the data from DAMPE experiment assuming nearby DM subhalo; for example, the data can be fitted with current DM thermal annihilation cross section $\langle \sigma v \rangle_0 = 3 \times 10^{-26}$cm$^3/$s, subhalo mass $M_{\rm sub}/M_{\bigodot} = 2.6 \times 10^8$, annihilation luminosity of subhalo $\mathcal{L}/$GeV$^2$cm$^{-3} = 1.0 \times 10^{66}$, and distance from the earth 0.3 kpc~\cite{Yuan:2017ysv} when we assume DM annihilate into only charged leptons by ratio of $e:\mu:\tau = 1:1:1$. The preferred astrophysical parameters depend on the value of $\langle \sigma v \rangle_0$ where we can obtain $\mathcal{O}(10^{-26})$cm$^3/$s to $\mathcal{O}(10^{-24})$cm$^3/$s around $m_{Z'} = 3$ TeV. Note that we prefer enhanced annihilation cross section since our DM annihilate into neutrinos, too. Thus the the mass relation $m_{Z'} \leq 2 M_X$ is relevant to get desired cross section as shown in the right plot of Fig.~\ref{fig:DM1}. 
}

 \begin{widetext}
\begin{center} 
\begin{table}[t]
\begin{footnotesize}
\begin{tabular}{|c||c|c|c|c|c|c|c|c|c|c|}\hline\hline  
 & \multicolumn{9}{c|}{Leptons} \\\hline
Fermions  ~&~ $L_{L_e}$ ~&~ $L_{L_\mu}$ ~&~ $L_{L_\tau}$ ~&~ $e_R$ ~&~ $\mu_R$ ~&~ $\tau_R$ ~&~ $L'_{e}$ ~&~ $L'_{\mu}$ ~&~ $L'_{\tau}$~
\\\hline 
$SU(3)_C$  & $\bm{1}$  & $\bm{1}$  & $\bm{1}$   & $\bm{1}$  & $\bm{1} $  & $\bm{1}$ & $\bm{1}$  & $\bm{1} $  & $\bm{1}$ \\\hline 
 $SU(2)_L$  & $\bm{2}$  & $\bm{2}$  & $\bm{2}$   & $\bm{1}$  & $\bm{1}$   & $\bm{1}$  & $\bm{2}$  & $\bm{2}$   & $\bm{2}$ \\\hline 
$U(1)_Y$  & $-\frac{1}{2}$ & $-\frac12$ & $-\frac12$  & $-1$ &  $-1$  &  $-1$  & $-\frac12$ &  $-\frac12$  &  $-\frac12$ \\\hline
 $U(1)_{e-\mu}$ & $1$  & $-1$ & $0$ & $1$  & $-1$   & $0$ & $1$  & $-1$   & $0$ \\\hline
$Z_2$  & $+$  & $+$ & $+$ & $+$ & $+$ & $+$& $-$ & $-$ & $-$ \\\hline
\end{tabular}
\caption{Field contents of fermions
and their charge assignments under $SU(2)_L\times U(1)_Y\times  U(1)_{e-\mu}\times Z_2$.}
\label{tab:2-1}
 \end{footnotesize}
\end{table}
\end{center}
\end{widetext}
\begin{table}[t]
\centering {\fontsize{10}{12}
\begin{tabular}{|c||c|c||c|c|c|}\hline\hline
&\multicolumn{2}{c||}{VEV$\neq 0$} & \multicolumn{2}{c|}{Inert } \\\hline
  Bosons  &~ $\Phi$  ~ &~ $\varphi$ ~ &~ $\Delta$   ~ &~ $S$ ~ \\\hline
$SU(2)_L$ & $\bm{2}$ & $\bm{1}$ & $\bm{3}$  & $\bm{1}$    \\\hline 
$U(1)_Y$ & $\frac12$ & $0$ & $1$  & $0$    \\\hline
 $U(1)_{e-\mu}$ & $0$  & $1$ & $0$ & $0$   \\\hline
$Z_2$ & $+$ & $+$  & $-$ & $-$ \\\hline
\end{tabular}%
} 
\caption{Field contents of bosons
and their charge assignments under $SU(2)_L\times U(1)_Y\times U(1)'\times Z_2$, where $SU(3)_C$ singlet for all bosons. }
\label{tab:2-2}
\end{table}

\subsection{Model 2: Scalar boson DM}
Here we consider the second model and discuss some phenomenologies in the case of boson DM candidate with  $U(1)_{e-\mu}$ gauged symmetry, where
original model has been discussed in ref.~\cite{Lee:2017ekw} in the framework of $U(1)_{\mu-\tau}$ symmetry.
We summarize our field contents and their field assignments in Table~\ref{tab:2-1} for fermion sector and Table~\ref{tab:2-2} for boson sector.
$Z_2$ odd parity plays a role in assuring the stability of our DM as usual, where we identify it to be gauge singlet inert boson $S$.
Although $S$ mixes with $\Delta$ that is also inert boson with isospin triplet, we expect its mixing is so small that we can neglect its mixing effect in the analysis of DM. 
Singlet $\varphi$ is the source of the spontaneous symmetry breaking of $U(1)_{e-\mu}$, whose VEV is denoted by $\langle\varphi\rangle\equiv v'/\sqrt2$.
{\it The Lagrangian of neutrino sector and valid Higgs potential} is then given by 
\begin{align}
-{\cal L}_{L} = & \sum_{\ell=e,\mu,\tau} \left[f_\ell\bar L_{L_\ell} L'_{R_\ell}S +M_\ell \bar L'_{L_\ell} L'_{R_\ell} \right]
+g_3 \bar L_{L_\tau}^C (i\sigma_2)\Delta L'_{L_\tau}  + g_2\bar L_{L_e}^C (i\sigma_2)\Delta L'_{L_\mu} 
+ g_3 \bar L_{L_\mu}^C (i\sigma_2)\Delta L'_{L_e} 
\nn\\&
+ y_{E_1} \varphi \bar L'_{L_e} L'_{R_\tau} +  y_{E_2} \varphi^* \bar L'_{L_\mu} L'_{R_\tau} 
-\lambda_0 \Phi^T(i\sigma_2)\Delta^\dag \Phi S
+ {\rm c.c.},
\label{eq:lag-quark}
\end{align}
where $L'\equiv [N,E]^T$.
We parametrize the scalar fields as 
\begin{align}
&\Phi =\left[
\begin{array}{c}
w^+\\
\frac{v+\phi+iz}{\sqrt2}
\end{array}\right],\ 
{\eta =\left[
\begin{array}{c}
\eta^+\\
\frac{\eta_R+i\eta_I}{\sqrt2}
\end{array}\right]} 
,\ 
\Delta =\left[
\begin{array}{cc}
\frac{\Delta^+}{\sqrt2} & \Delta^{++}\\
\Delta^{0} & -\frac{\Delta^+}{\sqrt2}
\end{array}\right],
\ \Delta_0=\frac{\Delta_R + i\Delta_I}{\sqrt2},\ \varphi=\frac{v'+\varphi_R+i z'}{\sqrt2},
\label{component}
\end{align}
where $v~\simeq 246$ GeV is VEV of the Higgs doublet, and $w^\pm$, $z$, and $z'$ are respectively GB 
which are absorbed by the longitudinal component of $W$, $Z$, and $Z'(\equiv Z_{e-\mu})$ boson.
Then we have two neutral  boson mass matrices $m^2_{R_1}$ and $m^2_{R_2}$ in the basis of $[\varphi_R,\phi]^T$ and $[S,\Delta_R]^T$, and these are diagonalized by $O_{R}^T m^2_{\varphi_R\phi}O_{R}\equiv$Diag[$m_{h_1},m_{h_2}$] and $O_{\alpha}^T m^2_{S\Delta}O_{\alpha}\equiv$Diag[$m_{H_1},m_{H_2}$] respectively, where the mixing source of $O_\alpha$ arises from the nontrivial quartic coupling $\lambda_0 \Phi^T(i\sigma_2)\Delta^\dag \Phi S$.
{For CP-even scalars, the lighter one is identified as the SM-like Higgs boson so that $m_{h_1} \simeq 125$ GeV, and we assume the mixing of $O_R$ is negligibly small for simplicity. In the following we just use $\varphi_R$ as the heavy CP-even scalar boson.}
The mass eigenstate of  inert bosons in basis of $[S,\Delta_R]^T$ is defined as
\begin{align}
&S=c_\alpha H_1+ s_\alpha H_2,\quad  \Delta_R=-s_\alpha H_1+ c_\alpha H_2,
\end{align}
where the mixing is written in terms of linear combinations of couplings of Higgs potential.

{We have heavy neutral gauge boson $Z'$ after $U(1)_{e - \mu}$ gauge symmetry breaking.
The mass of $Z'$ is given by
\begin{equation}
m_{Z'} = g' v',
\label{eq:MZp2}
\end{equation}
where $g'$ is the gauge coupling of $U(1)_{e - \mu}$.}

After the $e-\mu$ gauge symmetry breaking, {\it vector-like fermion mass matrix} can be written in the basis $[L'_{e},L'_{\mu},L'_{\tau}]^T$ as follows:
\begin{align}
M_{L'}\equiv \left[\begin{array}{ccc}  M_e & 0 & M_{e\tau}  \\ 0 & M_\mu & M_{\mu\tau} \\ M_{e\tau} & M_{\mu\tau} & M_\tau \end{array}\right],\label{eq:ML'}
\end{align}
where we {have simply assumed $M_{L'}$ to be a real symmetric matrix} and define $M_{e\tau}\equiv y_{E_1} v'/\sqrt2$ and $M_{\mu\tau}\equiv  y_{E_2} v'/\sqrt2$. Then $M_{L'}$ is diagonalized by orthogonal mixing matrix $V$ ($VV^T=1$) as 
\begin{align}
V^T M_{L'} V =D_N \equiv {\text{Diag.} }\left[M_1,M_2,M_3\right],\quad N_{{e,\mu,\tau}}=V N_{{1,2,3}},\label{eq:N-mix}
\end{align}
where $N_{1,2,3}(M_{1,2,3})$ is the mass eigenstate(eigenvalue).
 Then our active neutrino mass matrix is given at one-loop level and calculated as
\begin{align}
m_\nu^{th}&= 
g \epsilon V {D_N} R V^T f  + [g \epsilon V {D_N} R V^T f]^T ,
\label{eq:neut-mass}\\
\epsilon &\equiv
\left[\begin{array}{ccc}  0 & 1 & 0 \\ 1 & 0 & 0 \\ 0 & 0 & 1 \end{array}\right],\quad 
{ R} =\frac{s_\alpha c_\alpha}{(4\pi)^2}
\left[\frac{r_{k_2}\ln r_{k_2}}{1-r_{k_2}}-\frac{r_{k_1}\ln r_{k_1}}{1-r_{k_1}}\right],
\end{align}
where $r_{k_a} \equiv(m_{H_a}/M_k)^2$.
On the other hand, the neutrino mass matrix can be written in terms of experimental values as  $m_\nu^{exp}=U D^\nu U^T$.
Therefore, $m^{th}_\nu=m_\nu^{exp}$ should be satisfied.

It is worthwhile to mention the new contribution to the muon anomalous magnetic moment (muon $g-2$: $\Delta a_\mu$) that arises from $g$ with negative contribution and $f$  with positive contribution.
Since the experimental result is positively induced, we assume to be $g<<f$.
Then $(g-2)_\mu$ is given by 
\begin{align}
&\Delta a_\mu \approx
\frac{2 m_\mu^2 }{(4\pi)^2}\sum_{\ell=e,\mu,\tau}  F_{\mu\ell}  F^\dag_{\ell \mu}
\left[c_\alpha^2 F_2 [H_1,M_\ell] + s_\alpha^2 F_2 [H_2,M_\ell] \right] \nonumber \\
& \quad \quad \sim \frac{2 m_\mu^2 }{(4\pi)^2}\sum_{\ell=e,\mu,\tau}  F_{\mu\ell}  F^\dag_{\ell \mu} F_2 [S,M_\ell],\label{eq:g2}\\
& F_{2}(m_a,m_b)=\frac{2 m_a^6 +3 m_a^4 m_b^2 -6 m_a^2 m_b^4 +6 m_b^6+12 m_a^4 m_b^2 \ln(m_b/m_a)}{12(m_a^2-m_b^2)^4},
\end{align}
where $F\equiv fV$, and we have assumed to be $s_\alpha<<1$ that leads to $H_1\approx S$ in the last equation of (\ref{eq:g2}).
Since we have to consider the LFV constraints  such as $\mu\to e\gamma$, we impose the condition $\sum_{\ell=e,\mu,\tau}F_{e\ell}F^\dag_{\ell\mu}<<\sum_{\ell=e,\mu,\tau} F_{\mu\ell}F^\dag_{\ell\mu}$. Even in this case, one finds the sizable muon $(g-2)_\mu$.

\begin{figure}[t!]
\centering 
\includegraphics[width=8.4cm]{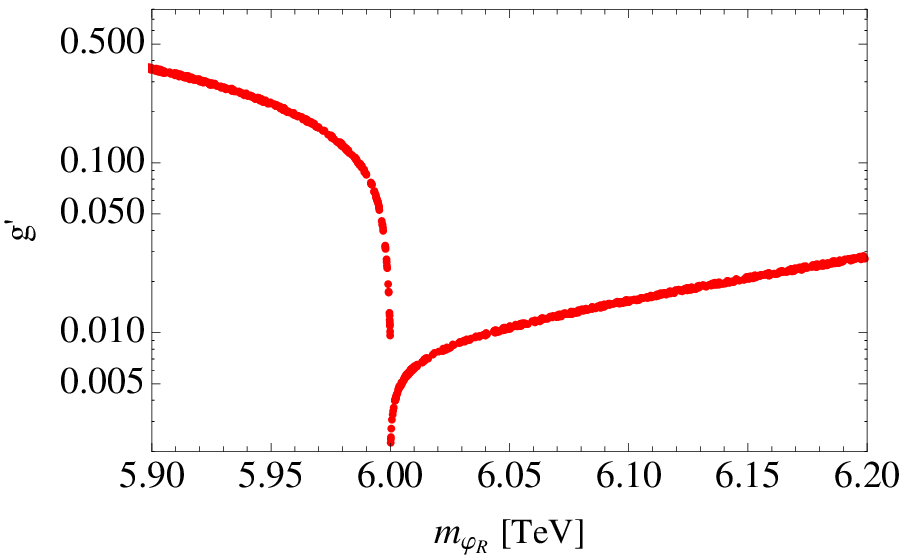}%
\includegraphics[width=8cm]{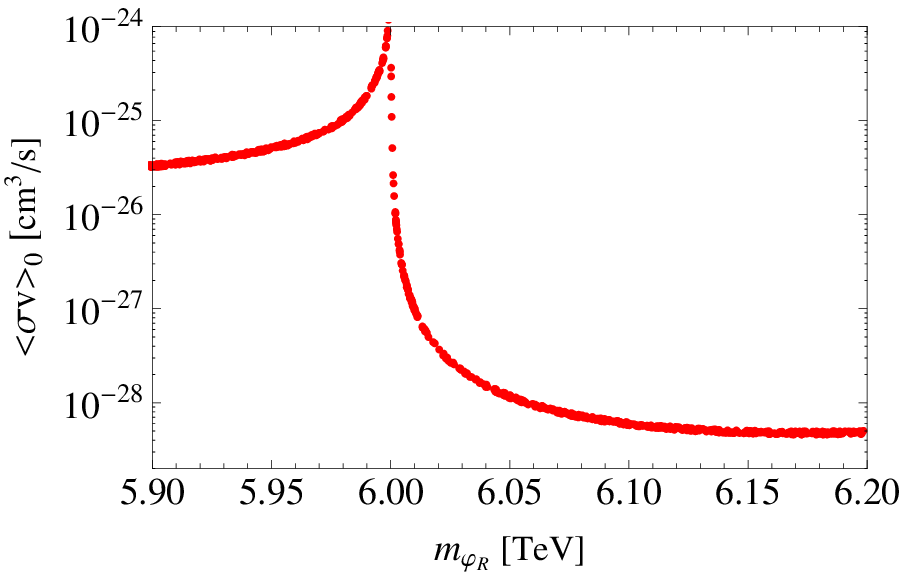}%
\caption{Left plot: The parameter points providing correct relic density on $\{m_{\varphi_R}, g' \}$ plane.  Right plot: Thermally averaged annihilation cross section at current Universe as a function of $m_{\varphi_R}$ using parameter points in the left plot.  Here the DM mass is fixed to be $M_X = 3.0$~TeV.}
\label{fig:DM2}
\end{figure}
{{\it Dark matter candidate inspired by DAMPE}:
Here we discuss possible explanation of the DAMPE excess by our scalar boson DM candidate as in the case of previous model.
Firstly we briefly formulate the valid interactions between scalar boson DM candidate ($X\equiv S,\ M_X\equiv m_S$) and the other particles
where we just neglect several  irrelevant interactions arising from Higgs potential and Yukawa couplings~\footnote{We have confirmed the cross section from the Yukawa coupling cannot be so large and relic density as well as DAMPE excess can be explained by these interactions.}.

The relevant interactions come from kinetic term and a part of Higgs potential, and the corresponding Lagrangian is given by
\begin{align}
{\cal L}\sim \mu X^2 \varphi_R + g'^2 v' Z'_\mu Z^{'\mu} \varphi_R
+ (\bar e \gamma^\mu e - \bar \mu \gamma^\mu \mu) Z'_\mu + (\bar \nu_e \gamma^\mu P_L \nu_e - \bar \nu_\mu \gamma^\mu P_L \nu_\mu) Z'_\mu,
\end{align}
where $\mu$ comes from a quartic coupling $S^2 \varphi^2$ in the Higgs potential and the mass of $Z'$ is given by Eq.~(\ref{eq:MZp2}).
The annihilation process is therefore four body modes: $2X\to \varphi_R\to 2Z'\to 2e(\mu)^\pm2e(\mu)^\mp(2\nu_e(\nu_\mu)^\pm2\nu_e(\nu_\mu)^\mp)$. This kind of process has also been analyzed by the group~\cite{Zu:2017dzm} and found a solution when $M_X\sim m_{\varphi_R}$ with $M_X\approx$ 3 TeV. This is a natural consequence that the excess should monochromatically be observed.   
Then we carry out parameter scan on $\{m_{H}, g_X \}$ space by fixing other parameters as $M_X = 3$ TeV, $m_{Z'} = 0.995 M_X$ and $\mu = 1$ TeV.
Note that we also implicitly apply LEP constraint for gauge coupling as in the case of previous model.
The left plot in Fig.~\ref{fig:DM2} shows the parameter points which can accommodate correct DM relic density as in the analysis of previous model. 
Then we show the current thermal annihilation cross section for the parameter points in the right plot in Fig.~\ref{fig:DM2}.
As in the previous model, we prefer slightly enhanced annihilation cross section as we have neutrino mode and $m_{\varphi_R} \leq 2 M_X$ is also suitable in explaining the DAMPE data.
}

\section{ Conclusions and discussions}
We have proposed two possibilities of the one-loop induced radiative seesaw models that link to the fermionic and bosonic dark matter candidates inspired by the DAMPE excess. In model 1, we have introduced gauged $U(1)_{e+\mu-\tau}$ and constructed the one-loop neutrino masses inside the Dirac type of DM, and found predictive two zero texture~$B_2$ that provides several predictions such that  inverted hierarchy is favored when the best fit observables are adapted, $m_{\nu_3}\approx \sqrt{\Delta m^2_{\rm atm}/(1-\cot^4\theta_{23})}$ and $m_{\nu_2}\approx m_{\nu_1}\approx m_{\nu_3}\cot^2\theta_{23}$ are derived at the leading order. The excess of DAMPE can be explained by the universal two body processes $\bar XX\to \ell(\nu_\ell)^\pm \ell(\nu_\ell)^\mp\ (\ell=e,\mu,\tau$) with s-channel, which is in good agreement with the recent experimental result of DAMPE at $M_X\approx$ 1.5 TeV. 

 In model 2, we have introduced gauged $U(1)_{e-\mu}$ and constructed the one-loop neutrino masses associated with the bosonic DM, and 
  can have explained the source of muon anomalous magnetic dipole moment, too.
The excess of DAMPE can be explained by the four body processes  $2X\to \varphi_R\to 2Z'\to 2e(\mu)^\pm2e(\mu)^\mp(2\nu_e(\nu_\mu)^\pm2\nu_e(\nu_\mu)^\mp)$ with s-channel, which is also in good agreement with the DAMPE result and there is a solution when $M_X\simeq m_{\varphi_R}$ with $M_X\simeq$ 3 TeV.


\section*{Acknowledgments}
\vspace{0.5cm}
H. O. would like thank Prof. Yue-Lin Sming for fruitful discussion about the DM candidate.



\begin{thebibliography}{99}

\bibitem{dampe}
J. Chang, Chinese Journal of Space Science 34, 550 (2014);
  J.~Chang {\it et al.} [DAMPE Collaboration],
  Astropart.\ Phys.\  {\bf 95}, 6 (2017)
  [arXiv:1706.08453 [astro-ph.IM]];

\bibitem{Ambrosi:2017wek} 
  G.~Ambrosi {\it et al.} [DAMPE Collaboration],
  doi:10.1038/nature24475
  arXiv:1711.10981 [astro-ph.HE].

\bibitem{Yuan:2017ysv} 
  Q.~Yuan {\it et al.},
  arXiv:1711.10989 [astro-ph.HE].

\bibitem{Adriani:2008zr} 
  O.~Adriani {\it et al.} [PAMELA Collaboration],
  Nature {\bf 458}, 607 (2009)
  [arXiv:0810.4995 [astro-ph]].
  
\bibitem{Accardo:2014lma} 
  L.~Accardo {\it et al.} [AMS Collaboration],
  Phys.\ Rev.\ Lett.\  {\bf 113}, 121101 (2014).



 
 
\bibitem{Fan:2017sor}
Y.~Z.~Fan, W.~C.~Huang, M.~Spinrath, Y.~L.~S.~Tsai and Q.~Yuan,
arXiv:1711.10995 [hep-ph].


\bibitem{Gu:2017gle} 
P.~H.~Gu and X.~G.~He,
arXiv:1711.11000 [hep-ph].

\bibitem{Duan:2017pkq} 
G.~H.~Duan, L.~Feng, F.~Wang, L.~Wu, J.~M.~Yang and R.~Zheng,
arXiv:1711.11012 [hep-ph].


\bibitem{Zu:2017dzm} 
L.~Zu, C.~Zhang, L.~Feng, Q.~Yuan and Y.~Z.~Fan,
arXiv:1711.11052 [hep-ph].


\bibitem{Tang:2017lfb} 
Y.~L.~Tang, L.~Wu, M.~Zhang and R.~Zheng,
arXiv:1711.11058 [hep-ph].


\bibitem{Chao:2017yjg} 
W.~Chao and Q.~Yuan,
arXiv:1711.11182 [hep-ph].


\bibitem{Gu:2017bdw} 
P.~H.~Gu,
arXiv:1711.11333 [hep-ph].


\bibitem{Athron:2017drj} 
P.~Athron, C.~Balazs, A.~Fowlie and Y.~Zhang,
arXiv:1711.11376 [hep-ph].


\bibitem{Cao:2017ydw} 
J.~Cao, L.~Feng, X.~Guo, L.~Shang, F.~Wang and P.~Wu,
arXiv:1711.11452 [hep-ph].


\bibitem{Duan:2017qwj} 
G.~H.~Duan, X.~G.~He, L.~Wu and J.~M.~Yang,
arXiv:1711.11563 [hep-ph].

\bibitem{Liu:2017rgs} 
X.~Liu and Z.~Liu,
arXiv:1711.11579 [hep-ph].

\bibitem{Chao-guo-li-shu} 
  W.~Chao, H.~K.~Guo, H.~L.~Li and J.~Shu,
  arXiv:1712.00037 [hep-ph].
  
\bibitem{Huang:2017egk} 
  X.~J.~Huang, Y.~L.~Wu, W.~H.~Zhang and Y.~F.~Zhou,
  arXiv:1712.00005 [astro-ph.HE].
  
\bibitem{Gao:2017pym} 
  Y.~Gao and Y.~Z.~Ma,
  arXiv:1712.00370 [astro-ph.HE].
  
\bibitem{Niu:2017hqe} 
  J.~S.~Niu, T.~Li, R.~Ding, B.~Zhu, H.~F.~Xue and Y.~Wang,
  arXiv:1712.00372 [astro-ph.HE].
  
  


\bibitem{Lee:2017ekw} 
  S.~Lee, T.~Nomura and H.~Okada,
  arXiv:1702.03733 [hep-ph].
  
  
\bibitem{Nomura:2017xko} 
  T.~Nomura and H.~Okada,
  arXiv:1706.01321 [hep-ph].

\bibitem{Nomura:2017tzj} 
  T.~Nomura and H.~Okada,
  arXiv:1706.05268 [hep-ph].

\bibitem{Nomura:2017psk} 
  T.~Nomura and H.~Okada,
  arXiv:1707.06083 [hep-ph].

\bibitem{Nomura:2017lsn} 
  T.~Nomura and H.~Okada,
  arXiv:1710.10028 [hep-ph].

\bibitem{Ma:2006km} 
  E.~Ma,
  Phys.\ Rev.\ D {\bf 73}, 077301 (2006)
  [hep-ph/0601225].

\bibitem{Fritzsch:2011qv} 
  H.~Fritzsch, Z.~z.~Xing and S.~Zhou,
  JHEP {\bf 1109}, 083 (2011)
  [arXiv:1108.4534 [hep-ph]].
  
\bibitem{Baek:2015mna} 
  S.~Baek, H.~Okada and K.~Yagyu,
  JHEP {\bf 1504}, 049 (2015)
  [arXiv:1501.01530 [hep-ph]].
  
\bibitem{pdg}
K.A. Olive et al. (Particle Data Group), Chin. Phys. C, 38, 090001 (2014) 
and 2015 update.
  
\bibitem{TheMEG:2016wtm} 
  A.~M.~Baldini {\it et al.} [MEG Collaboration],
  Eur.\ Phys.\ J.\ C {\bf 76}, no. 8, 434 (2016).
  
       \bibitem{Belanger:2014vza} 
  G.~Belanger, F.~Boudjema, A.~Pukhov and A.~Semenov,
  arXiv:1407.6129 [hep-ph].
 
\bibitem{Ade:2015xua} 
  P.~A.~R.~Ade {\it et al.} [Planck Collaboration],
  Astron.\ Astrophys.\  {\bf 594}, A13 (2016)
  [arXiv:1502.01589 [astro-ph.CO]].
  
  
\bibitem{Schael:2013ita} 
  S.~Schael {\it et al.} [ALEPH and DELPHI and L3 and OPAL and LEP Electroweak Collaborations],
  Phys.\ Rept.\  {\bf 532}, 119 (2013)
  [arXiv:1302.3415 [hep-ex]].
  
\end{thebibliography}
\end{document}